\documentclass[11pt]{article}

\usepackage{natbib}
\usepackage{hyperref}
\usepackage{amsfonts,amstext,amsmath,amsthm}
\usepackage{tikz}
\usepackage[labelfont=bf, font=small]{caption}
\usepackage{fullpage}
\setcitestyle{authoryear,open={(},close={)}}
\newcommand*{\doi}[1]{\href{http://dx.doi.org/#1}{doi:#1}}

\usepackage{algorithm}
\usepackage{algpseudocode}

\usepackage{mathtools}
\usepackage{dsfont}
\usepackage{booktabs}
\usepackage{fontawesome}
\usepackage{listofitems} 

\definecolor{ForestGreen}{HTML}{228B22}
\definecolor{OptimalBlue}{HTML}{1E88E5}

\newlength{\figwidth}
\newlength{\tabwidth}
\newcommand{\E}{\mathbb{E}}

\DeclareMathOperator{\var}{Var}

\newcommand{\indep}{\perp\nolinebreak\hspace{-5pt}\perp} 

\newcommand{\given}{\,|\,}
\newcommand{\RR}{\mathbb{R}}
\newcommand{\comet}{\texttt{comets}}
\newcommand{\pkg}[1]{\texttt{#1}}
\newcommand{\proglang}[1]{\textsf{#1}}

\renewcommand{\epsilon}{\varepsilon}
\renewcommand{\hat}{\widehat}

\DeclareMathOperator{\RNA}{RNAseq}
\DeclareMathOperator{\miRNA}{miRNA}
\DeclareMathOperator{\DNA}{DNAm}

\usetikzlibrary{positioning, calc, shapes.geometric, shapes.multipart,
  shapes, arrows.meta, arrows, quotes,
  decorations.markings, external, trees}
\tikzstyle{line} = [draw, -latex']
\tikzstyle{Arrow} = [
        thick,
        decoration={
                markings,
                mark=at position 1 with {
                        \arrow[thick]{latex}
} },
        shorten >= 3pt, preaction = {decorate}
        ]
        
\graphicspath{{Fig/}}

\begin{document}

\title{\bf Algorithm-agnostic significance testing in supervised learning with 
multimodal data}

\author{Lucas Kook\textsuperscript{1*}, Anton Rask Lundborg\textsuperscript{2}}

\date{\small 
\textsuperscript{1}Institute for Statistics and Mathematics, Vienna University
of Economics and Business, Austria\\
  \textsuperscript{2}Department of Mathematical Sciences, University of
  Copenhagen, Denmark\\
\textsuperscript{*}E-mail:~\texttt{lucasheinrich.kook@gmail.com}}

\maketitle

\begin{abstract}
Valid statistical inference is crucial for decision-making but difficult to
obtain in supervised learning with multimodal data, e.g.,\ combinations of
clinical features, genomic data, and medical images. Multimodal data often
warrants the use of black-box algorithms, for instance, random forests or neural
networks, which impede the use of traditional variable significance tests.
We address this problem by proposing the use of COvariance Measure Tests
(COMETs), which are calibrated and powerful tests that can be combined with any
sufficiently predictive supervised learning algorithm. We apply COMETs to
several high-dimensional, multimodal data sets to illustrate (i) variable
significance testing for finding relevant mutations modulating drug-activity,
(ii) modality selection for predicting survival in liver cancer patients with
multiomics data, and (iii) modality selection with clinical features and medical
imaging data. In all applications, COMETs yield results consistent with domain
knowledge without requiring data-driven pre-processing which may invalidate
type~I error control. These novel applications with high-dimensional multimodal
data corroborate prior results on the power and robustness of COMETs for
significance testing.
COMETs are implemented in the \comet{} \proglang{R} package available on CRAN
and \pkg{pycomets} \proglang{Python} library available on GitHub.
Source code for reproducing all results is available at
\url{https://github.com/LucasKook/comets}. All data sets used in this work are
openly available.
\end{abstract}

\section{Introduction}\label{sec:intro}

A fundamental challenge of modern bioinformatics is dealing with the 
increasingly multimodal nature of data 
\citep{cheerla2019multimodal,khandakar2021multiomicsgan,stahlschmidt2022multimodal}. 
The task of \emph{supervised learning}, that is, the problem of predicting a
response variable $Y$ from features $X$, has received considerable attention in
recent years resulting in a plethora of algorithms for a wide range of settings 
that permit prediction using several data modalities simultaneously 
\citep{hastie2009elements}. With the advent of deep learning, even non-tabular 
data modalities, such as text or image data, can be included without requiring 
manual feature engineering \citep{lecun2015deep}.
Methods such as these are highly regularized (if trained correctly) which 
minimizes the statistical price of adding too many irrelevant variables. 
However, continuing to collect features or modalities that do not contribute 
to the predictiveness of a model still has an economic cost and, perhaps more
importantly, it is of scientific interest to determine whether a particular
feature or modality $X$ adds predictive power in the presence of 
additional features or modalities $Z$ \citep{smucler2022note}. 

The problem of determining which features or modalities are significantly
associated with the response is usually addressed by means of \emph{conditional
independence testing}. The response $Y$ is independent of the modality $X$
given further modalities $Z$ if the probability that $Y$
takes any particular
value knowing both $X$ and $Z$ is the same as the probability knowing just 
$Z$. In particular, $X$ does not help in predicting $Y$ 
if $Z$ is taken into account already
(see Section~\ref{sec:ci-testing} for a more precise definition).

Traditional variable significance tests start by posing a parametric 
relationship between the response $Y$ and features $X$ and $Z$,
for instance, the Wald test in a generalized linear model. 
When $X$ or $Z$ are complicated data modalities, it is seldom possible to 
write down a realistic model for their relationship with $Y$; 
thus a different approach is required. 
Furthermore, even when models can be explicitly parametrized, it is not clear
that the resulting tests remain valid when the model is not specified correctly 
\citep{shah2023double}. 

More recently, kernel-based conditional independence tests have been proposed
which use a characterization of conditional independence by means of kernel
embeddings to construct tests \citep{zhang2012kernel,strobl2019approximate}.
However, these tests are difficult to calibrate in practice and rely intimately
on kernel ridge regression.
Several alternative algorithm-agnostic tests have been developed under the 
so-called `Model-X' assumption where one supposes that a model is known (or at 
least estimable to high precision) for the full distribution of $X$ given 
$Z$ \citep{candes2016panning,berrett2019cond}. Given the difficulty of learning
conditional distributions such an assumption is rarely tenable. Algorithm-agnostic
variable importance measures have also been developed with statistically optimal
estimators \citep{williamson2021nonparametric,williamson2023general}. However,
efficient estimation of an importance measure does not necessarily translate to 
an optimal test to distinguish between conditional dependence and independence 
\citep[see, e.g.,\ the introduction of][]{lundborg2022projected}.

\begin{figure*}[t!]
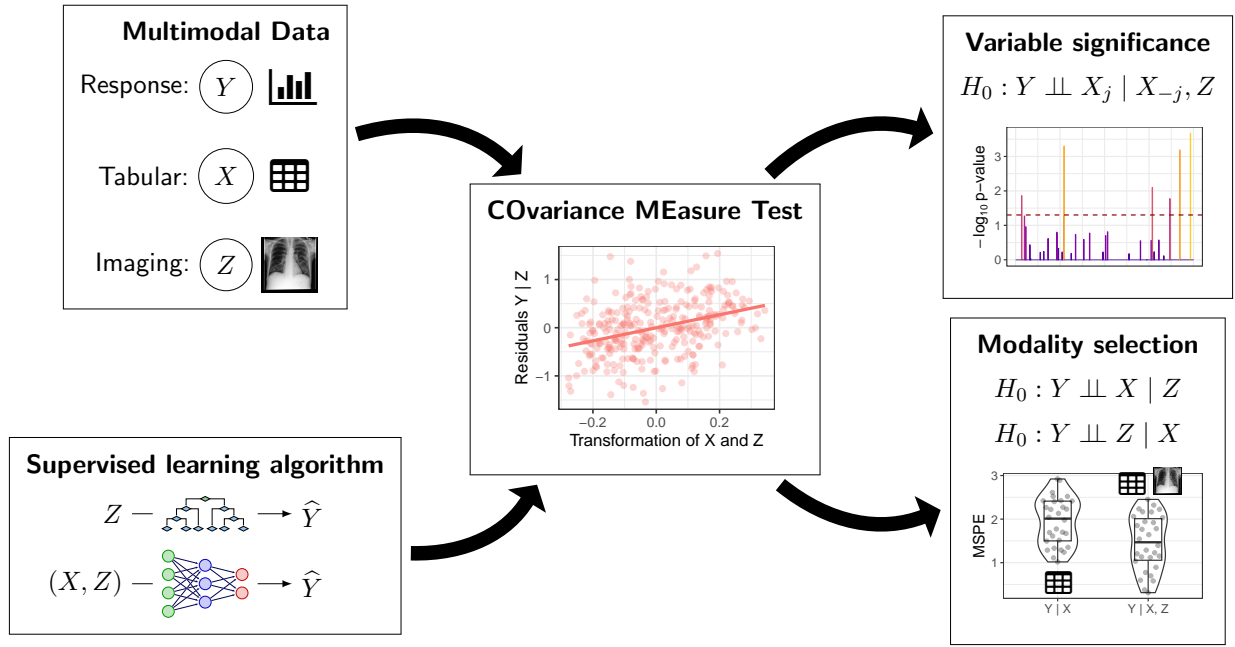

    \centering
    \resizebox{\textwidth}{!}{%
\begin{tikzpicture}[auto, Arr/.style={-{Latex[length=1.5mm]}}]
\node (data) {\input{Fig/data}};
\node[below=1.58cm of data] (sla) {\fbox{%
\begin{tikzpicture}[auto, Arr/.style={-{Latex[length=1.5mm]}}]
  \usetikzlibrary{shapes,arrows,fit,calc,positioning}
  \tikzset{box/.style={draw, diamond, thick, text centered, minimum height=0.5cm, minimum width=1cm}}
  \tikzset{line/.style={draw, thick, -latex'}}
\node (sla) {\textbf{\textsf{Supervised learning algorithm}}};
\node[below=0cm of sla] (rf) {\resizebox{1.25cm}{!}{%
\begin{tikzpicture}[auto]
\node [box, fill=ForestGreen!50]                                    (x3)      {};
\node [box, below=0.5cm of x3, xshift=-2cm, fill=OptimalBlue!50]    (x1sx)    {};
\node [box, below=0.5cm of x3, xshift=2cm, fill=OptimalBlue!50]     (x2dx)    {};
\node [box, below=0.5cm of x1sx, xshift=-1cm, fill=OptimalBlue!50]  (x2sx)    {};
\node [box, below=0.5cm of x2sx, xshift=1cm, fill=OptimalBlue!50]   (A2sx)    {};
\node [box, below=0.5cm of x2sx, xshift=-1cm, fill=OptimalBlue!50]  (A1sx)    {};
\node [box, right=0.25cm of A2sx, fill=OptimalBlue!50]                 (A3sx)    {};

\node [box, below=0.5cm of x2dx, xshift=1cm, fill=OptimalBlue!50]   (x1dx)    {};
\node [box, below=0.5cm of x1dx, xshift=-1cm, fill=OptimalBlue!50]  (A2dx)    {};
\node [box, below=0.5cm of x1dx, xshift=1cm, fill=OptimalBlue!50]   (A3dx)    {};
\node [box, left=0.25cm of A2dx, fill=OptimalBlue!50]                (A1dx)    {};
\path [line] (x3) -|         (x2dx);
\path [line] (x3) -|         (x1sx);
\path [line] (x2dx) -|       (x1dx);
\path [line] (x2dx) -|       (A1dx);
\path [line] (x1dx) -|       (A2dx);
\path [line] (x1dx) -|       (A3dx);
\path [line] (x1sx) -|       (x2sx);
\path [line] (x1sx) -|       (A3sx);
\path [line] (x2sx) -|       (A1sx);
\path [line] (x2sx) -|       (A2sx);
\end{tikzpicture}}};
\node[left=0.3cm of rf] (inp) {$Z$};
\node[right=0.5cm of rf] (oup) {$\hat{Y}$};
\draw[-] (inp) -- (rf);
\draw[-latex] (rf) -- (oup);
\node[below=0cm of rf] (nn) {\resizebox{1.25cm}{!}{
\begin{tikzpicture}[x=2.4cm,y=1.2cm]
\tikzset{>=latex} 
\colorlet{myred}{red!80!black}
\colorlet{myblue}{blue!80!black}
\colorlet{mygreen}{green!60!black}
\colorlet{mydarkred}{myred!40!black}
\colorlet{mydarkblue}{myblue!40!black}
\colorlet{mydarkgreen}{mygreen!40!black}
\tikzstyle{node}=[very thick,circle,draw=myblue,minimum size=22,inner sep=0.5,outer sep=0.6]
\tikzstyle{connect}=[->,thick,mydarkblue,shorten >=1]
\tikzset{ 
  node 1/.style={node,mydarkgreen,draw=mygreen,fill=mygreen!25},
  node 2/.style={node,mydarkblue,draw=myblue,fill=myblue!20},
  node 3/.style={node,mydarkred,draw=myred,fill=myred!20},
}
\def\nstyle{int(\lay<\Nnodlen?min(2,\lay):3)} 

  \readlist\Nnod{4,3,2} 
  \readlist\Nstr{n,m,k} 
  \readlist\Cstr{x,h^{(\prev)},y} 
  \def\yshift{0} 
  
  \foreachitem \N \in \Nnod{
    \def\lay{\Ncnt} 
    \pgfmathsetmacro\prev{int(\Ncnt-1)} 
    \foreach \i [evaluate={\c=int(\i==\N); \y=\N/2-\i-\c*\yshift;
                 \x=\lay; \n=\nstyle;
                 \index=(\i<\N?int(\i):"\Nstr[\n]");}] in {1,...,\N}{ 
      \node[node \n] (N\lay-\i) at (\x,\y) {};
      
      \ifnumcomp{\lay}{>}{1}{ 
        \foreach \j in {1,...,\Nnod[\prev]}{ 
          \draw[white,line width=1.2,shorten >=1] (N\prev-\j) -- (N\lay-\i);
          \draw[connect] (N\prev-\j) -- (N\lay-\i);
        }
        \ifnum \lay=\Nnodlen
        \fi
      }{
      }
      
    }
    \path (N\lay-\N) --++ (0,1+\yshift) node[midway,scale=1.6] {}; 
  }
  
  
\end{tikzpicture}}};
\node[left=0.3cm of nn] (inp2) {$(X, Z)$};
\node[right=0.5cm of nn] (oup2) {$\hat{Y}$};
\draw[-] (inp2) -- (nn);
\draw[-latex] (nn) -- (oup2);
\end{tikzpicture}%
}};
\node[right=1.5cm of data, yshift=-2.5cm] (gcm) {\input{Fig/gcm}};
\node[right=1.5cm of gcm, yshift=2.5cm] (varsig) {\input{Fig/varsig}};
\node[below= 0cm of varsig] (modsec) {\input{Fig/modsec}};
\draw[-{Triangle[width=18pt,length=8pt]}, line width=6pt] (data) edge[bend left] (gcm);
\draw[-{Triangle[width=18pt,length=8pt]}, line width=6pt] (sla) edge[bend right] (gcm);
\draw[-{Triangle[width=18pt,length=8pt]}, line width=6pt] (gcm) edge[bend left] (varsig);
\draw[-{Triangle[width=18pt,length=8pt]}, line width=6pt] (gcm) edge[bend right] (modsec);
\end{tikzpicture}%
}
    \caption{%
    Overview of the proposed algorithm-agnostic significance testing framework 
    for multimodal data using COMETs.
    \textbf{Variable significance}: Differential gene expression can be assessed 
    in presence of the potentially high-dimensional/non-tabular confounder $Z$.
    \textbf{Modality selection}: Entire modalities can be subjected to significance 
    testing, which lends itself to modality selection in multi-omics applications.
    MSPE: Mean squared prediction error.
    }\label{fig:overview}
\end{figure*}

In this paper, we describe a family of significance tests referred to collectively
as COvariance Measure Tests (COMETs) which are algorithm-agnostic and valid (in
the sense of controlling the probability of false positives) as long as the
algorithms employed are sufficiently predictive. We will primarily focus on the 
Generalised Covariance Measure (GCM) test \citep{shah2020hardness} which we think 
of as an `all-purpose' test that should be well-behaved in most scenarios and 
the more complicated Projected Covariance Measure (PCM) test which is more 
flexible but may require a more careful choice of algorithms.
Figure~\ref{fig:overview} gives an overview of the proposed algorithm-agnostic
significance testing framework based on COMETs and the types of applications
that are presented in this manuscript. 
The main contribution of this work is to illustrate the use of the GCM and PCM
test in the context of multimodal, non-tabular data.

\section{Methods}

In this section, we first provide some background on conditional independence.
We then move on to describe the computation of the GCM and PCM tests
in addition to the assumptions required for their validity.
Finally, we describe the datasets which we will analyze in 
Section~\ref{sec:results}.

\subsection{Background on conditional independence}
\label{sec:ci-testing}

For a real-valued response $Y$ and features $X$ and $Z$, we say that $Y$ is 
\emph{conditionally independent of $X$ given $Z$} and write $Y \indep X \given Z$ if
\begin{equation}
\label{eq:ci}
\E[h(Y) \given X, Z] = \E[h(Y) \given Z] \quad \text{for all functions $h$.} 
\end{equation}
That is, for any transformation $h$ of $Y$, the best predictor (in a 
mean-squared error sense) of $h(Y)$ using both $X$ and $Z$ is equal to the 
best predictor using just $Z$.\footnote{An alternative characterization, when
$Y$ has a conditional density given $X$ and $Z$ denoted by $f_{Y|X,Z}(y | x, z)$,
is given by: $f_{Y|X,Z}(y|x,z) = f_{Y|Z}(y|z)$ if and only if $Y$ is independent 
of $X$ given $Z$.} 

A helpful starting point for the construction of a conditional independence test
is to consider the product of a population residual from a $Y$ on $Z$ regression 
$\varepsilon := Y- \E[Y \given Z]$ and, for now considering a one-dimensional $X$,
from an $X$ on $Z$ regression $\xi := X - \E[X \given Z]$. 
As these are population residuals, $Z$ is no longer helpful
in predicting their values, so $\E[\varepsilon \given Z] = \E[\varepsilon] = 0$
and similarly $\E[\xi\given Z] =0$. When $Y \indep X \given Z$, we can say more:
the product of the residuals is also mean zero since
\begin{align}
\label{eq:residual-product}
\E[\varepsilon \xi ] = \E[\E[\varepsilon \xi  \given X, Z]] = \E[\E[\varepsilon   
\given X, Z] \xi ] = \E[\E[\varepsilon   \given Z] \xi ] = 0,
\end{align}
where the second equality uses that 
$\xi$ is perfectly predicted using $X$ and $Z$,
the third equality uses \eqref{eq:ci} with $h(y) = y$
and the final equality uses 
$\E[\varepsilon \given Z] = 0$. 
The GCM test is based on testing whether $\E[\epsilon\xi] = 0$
and we will describe the details of how to compute it in 
Section~\ref{sec:computation-tests}.
For the GCM test to perform well,
it is important to determine when we can expect $\E[\epsilon\xi]$
to be non-zero under conditional dependence. When $Y$ follows
a partially linear model given $X$ and $Z$, that is, $\E[Y \given X, Z] = \theta X 
+ h(Z)$ for some function $h$, then $\E[\epsilon\xi] \neq 0$ exactly when 
$\theta \neq 0$ and the magnitude of $\E[\epsilon\xi]$ is proportional to 
$\theta$. This includes as a special case the linear model for $Y$ given $X$ and $Z$. 
There is a natural generalization of \eqref{eq:residual-product} to the case
where $X$ is a $d$-dimensional vector,
where the equation is interpreted 
component-wise in $X$. Although the GCM is also defined in
these settings, computing the test involves many regressions 
when $X$ is high-dimensional which can be impractical (see Section~\ref{sec:comparison}).

\begin{figure*}[!ht]
    \centering
    \resizebox{\textwidth}{!}{%
    \input{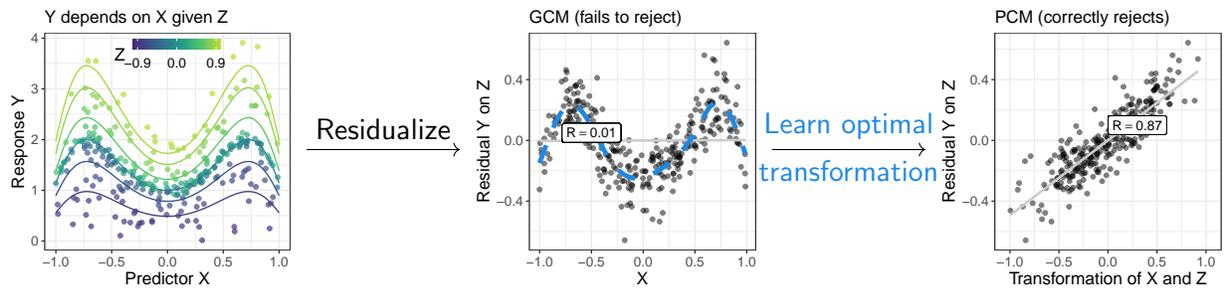}}
    \caption{%
    Illustration of the GCM and PCM test under the alternative that $Y \in \RR$
    is not conditionally independent of $X \in \RR$ given $Z \in \RR$, where
    $Y \coloneqq f(X) g(Z) + \varepsilon$ and $f(x) := 1 + \sin(3x^2)$, 
    $g(z) := 1 + z^3$. The GCM test first computes the
    residual for the regression of $Y$ on $Z$, which shows no correlation with
    $X$. Thus, the GCM fails to reject (correlation coefficient $R = 0.01$). 
    The PCM, in addition, learns the optimal
    transformation of $X$ (depending on $Z$) to test conditional mean independence 
    of $Y$ and $X$ given $Z$. Thus, in this example, the PCM test correctly 
    rejects ($R = 0.87$). Although the residuals in the second panel are clearly
    not independent,
    it is not valid to conclude conditional dependence from rejecting an independence 
    test here \protect{\citep[see][Example~1]{shah2020hardness}}.
    }\label{fig:gcmpcm}
\end{figure*}

Unfortunately, it is not difficult to come up with examples where $Y \not\indep X 
\given Z$ but $\E[\epsilon\xi] = 0$. For instance, if $X$ and $Z$ are 
independent and standard normally distributed and $\E[Y \given X, Z] = X^2$, then
$\E[Y \given Z] = 1$ (since $Z$ carries no information about $X$ so the best predictor
is just the mean of $Y$) hence 
\[
\E[\epsilon\xi] = \E[(X^2 - 1) X] = 0,
\]
using that $\E[X] = \E[X^3] = 0$ for a standard normal variable.
A more elaborate example is given in Figure~\ref{fig:gcmpcm} 
(left and middle panel) and even more examples exist when $X$ and 
$Z$ are dependent (see \citealt{lundborg2022projected}, Section~6 and 
\citealt{scheidegger2022weighted}, Section~3.1.2).
We now describe a test which can detect such dependencies.

A more ambitious target is to detect whenever an arbitrary 
(e.g.,\ non-tabular)
$X$ is helpful for the prediction of $Y$ in the presence of $Z$ measured in 
terms of mean-squared error. To achieve this goal, we can use the fact, derived 
in the same way as \eqref{eq:residual-product}, that 
\[
\E[\varepsilon f(X, Z)] = 0 \quad \text{for all functions $f$},
\]
whenever $Y \indep X \given Z$. The GCM targets the quantity involving the 
function $f(X, Z) = X$. However, by instead using
$f(X, Z) = \E[Y \given X, Z] - \E[Y \given Z]$ (which depends on the joint 
distribution of $X$ and $Z$), we obtain that 
\begin{equation}\label{eq:tau}
\E[\varepsilon f(X, Z)] = \E[(\E[Y \given Z] - \E[Y \given X, Z])^2] =: \tau.
\end{equation}
This quantity is strictly greater than $0$ if and only if $X$ is helpful for 
the prediction of $Y$ in the presence of $Z$. The PCM test is based on testing 
whether $\tau = 0$ and we will describe the details of how to compute it in 
Section~\ref{sec:computation-tests}. In fact, the PCM is based on an alternative
$f$ given by $f(X, Z) =  (\E[Y \given X, Z] - \E[Y \given Z])/\var(Y \given X, Z)$
that turns out to result in a more powerful test 
\citep[see Figure~\ref{fig:gcmpcm}  and][Section~1.1]{lundborg2022projected}.
An added benefit of tests targeting $\tau$ is that no regressions are needed with 
$X$ as the response, which can vastly reduce the computational burden when compared 
to tests that target $\E[\varepsilon \xi]$.

The targets mentioned above rely intimately on population quantities that are unknown 
and hence need to be estimated when computing tests in practice. To ensure that the
estimation errors do not interfere with the performance of the tests, we need to be
able to learn the functions to a sufficient degree of accuracy. These requirements put
restrictions on when the GCM and PCM are valid tests but such restrictions are not
unique to these tests.
In fact, unless $Z$ is discrete, it is impossible to construct
an assumption-free conditional independence test that simultaneously controls the
probability of false rejections and is able to detect dependence
\citep{shah2020hardness,kim2022local}. 
This result implies that additional assumptions need to be imposed to
ensure the feasibility of testing for conditional independence.

\subsection{Covariance measure tests}
\label{sec:computation-tests}
We now describe the specifics of computing the GCM and the PCM. For the remainder 
of this section, we assume that we have a dataset consisting of $n$ independent 
observations of a real-valued response $Y$ and some additional features 
or modalities $X$ and $Z$.

\subsubsection{Generalised Covariance Measure (GCM) test}
The GCM test is based on \eqref{eq:residual-product} but to compute the test in 
practice, we need to form an empirical version of the equation. For simplicity,
we consider, for now, $X \in \mathbb{R}$. 
Let $\widehat{\xi}_i$ denote the residual
for the $i$th observation from regressing $X$ on $Z$ and similarly 
$\widehat{\varepsilon}_i$ from regressing $Y$ on $Z$. We now test $H_0 : 
\E[\varepsilon \xi]  = 0$ by comparing
\begin{equation}
\label{eq:T}
T := \frac{\left(\frac{1}{\sqrt{n}} \sum_{i=1}^n \widehat{\varepsilon}_i
\widehat{\xi}_i\right)^2}{\frac{1}{n} 
\sum_{i=1}^n \widehat{\varepsilon}_i^2\widehat{\xi}_i^2 - \left(\frac{1}{n} 
\sum_{i=1}^n \widehat{\varepsilon}_i\widehat{\xi}_i\right)^2}
\end{equation}
to a $\chi^2_1$ distribution. The term inside the square in the numerator is 
$\sqrt{n}$ times an estimate of \eqref{eq:residual-product} while the
denominator standardizes the variance of the test statistic. The test statistic 
in \eqref{eq:T} is approximately $\chi^2_1$ for large enough sample sizes if the
regression methods employed are sufficiently predictive
and $Y \indep X \given Z$
\citep[][Theorem 6]{shah2020hardness}. Note that the procedure above did not use
anything special about $Z$ other than the existence of a regression method that
can approximate the conditional expectations of $Y$ and $X$ given $Z$. The
computations above naturally generalize to settings where $X \in \mathbb{R}^d$ 
and we summarize the general procedure in Algorithm~\ref{alg:gcm}.

\begin{algorithm}
\caption{GCM test of $H_0 : \E[\epsilon\xi] = 0$}\label{alg:gcm}
\begin{algorithmic}[1]
\Require Data set containing $n$ realizations of $Y \in \RR$, $X \in\RR^d$, 
$Z \in \mathcal{Z}$, algorithms for the regressions of $Y$ on $Z$ and $X_j$ on 
$Z$ for $j \in \{1, \dots, d\}$
\State Run the regression of $Y$ on $Z$ and compute residuals $\hat\epsilon_i$
\State Run the regressions of $X_j$ on $Z$ for $j \in \{1, \dots, d\}$ \newline
and compute residuals $\hat\xi_i \gets 
(\hat\xi_{i1}, \dots, \hat\xi_{id})$
\State Compute the mean residual product:
$L \gets \frac{1}{n} \sum_{i=1}^n \hat\epsilon_i\hat\xi_i$
\State Compute the residual variance:
$\Sigma \gets \frac{1}{n} \sum_{i=1}^n \hat\epsilon_i^2\hat\xi_i\hat\xi_i^\top
- LL^\top$
\State Compute the test statistic:
$T \gets n \lVert \Sigma^{-1/2} L\rVert_2^2$
\State Compute the $p$-value: $p \gets 1 - F_{\chi^2_d}(T)$
\State \Return{$p$}
\end{algorithmic}
\end{algorithm}

\subsubsection{Projected Covariance Measure (PCM) test}\label{sec:pcm}

The computation of the PCM test is more challenging than the computation of the GCM 
test since the PCM requires learning $f(X, Z) = (\E[Y \given X, Z] - \E[Y \given Z])
/\var(Y \given X, Z)$ to be able to estimate $\E[\varepsilon f(X, Z)]$. Furthermore,
$f$ cannot be learned on the same observations that are used to compute the test 
statistic as this would potentially result in dependence between the residuals
constituting the test statistic and thus in many false rejections when $Y \indep X
\given Z$.

The first step when computing the test statistic of the PCM test is therefore to split
the dataset in two halves $D_1$ and $D_2$ of equal size
(for simplicity, we assume that we have $2n$ observations,
so both $D_1$ and $D_2$ are of size $n$). 
On $D_2$, we compute an estimate $\widehat{f}$
of $f$ by first regressing $Y$ on $X$ and $Z$ yielding an estimate $\widehat{g}$ and 
regressing $Y$ on $Z$ yielding an estimate $\widehat{m}$. We then regress $(Y - 
\widehat{g}(X, Z))^2$ on $(X, Z)$ on $D_2$ yielding an estimate of $\var(Y \given 
X, Z)$ which we denote $\widehat{v}$. We now set $\widehat{f}(x, z) := (\widehat{g}(x,
z) - \widehat{m}(z))/\widehat{v}(x, z)$ and, working on $D_1$, we regress $Y$ on $Z$
yielding a residual for the $i$th observation $\widehat{\varepsilon}_i$ and we regress
$\widehat{f}(X, Z)$ on $Z$ yielding a residual $\widehat{\zeta}_i$. Finally, we compute
\begin{equation}
\label{eq:T-pcm}
T := \frac{\frac{1}{\sqrt{n}} \sum_{i=1}^n \widehat{\varepsilon}_i\widehat{\zeta}_i}{
\left(\frac{1}{n} \sum_{i=1}^n \widehat{\varepsilon}_i^2\widehat{\zeta}_i^2 - 
\left(\frac{1}{n} \sum_{i=1}^n \widehat{\varepsilon}_i\widehat{\zeta}_i\right)^2 \right)^{1/2}}
\end{equation}
and reject the null by comparing to a standard normal distribution. In fact, as the 
target of $T$ in \eqref{eq:tau} is positive under conditional dependence,
we perform a one-sided test
which rejects when $T$ is large. The test statistic in \eqref{eq:T-pcm} is
approximately standard Gaussian if the regression methods employed for the 
$\widehat{f}(X, Z)$ on $Z$ and $Y$ on $Z$ are sufficiently predictive,
the estimates $\widehat{f}$ 
are not too complicated
and $Y \indep X \given Z$ 
\citep[][Theorem 4]{lundborg2022projected}. The test is powerful against alternatives
where $\widehat{f}$ is correlated with the true $f$ and the aforementioned regression 
methods remain powerful \citep[][Theorem 5]{lundborg2022projected}. We summarize the 
procedure in Algorithm~\ref{alg:pcm} below.\footnote{In this description and in 
Algorithm~\ref{alg:pcm} we have omitted a few minor corrections to the estimation of
$\widehat{f}$ that are done for numerical stability or as finite sample corrections.
The full version of the algorithm with these additions
is given in \citet[][Algorithm~1]{lundborg2022projected}.}

\begin{algorithm}
\caption{PCM test of $H_0 : \E[Y \given X, Z] = \E[Y \given Z]$}\label{alg:pcm}
\begin{algorithmic}[1]
\Require Data containing $2n$ realizations of $Y \in \RR$, $X \in \mathcal{X}$,
$Z \in \mathcal{Z}$, regression algorithms as outlined in Section~\ref{sec:pcm}
\State Split the data randomly into equal parts $D_1$ and $D_2$ of size $n$
\State Regress $Y$ on $X, Z$ using $D_2$ to obtain $\hat{g}$
\State Regress of $Y$ on $Z$ using $D_2$ to obtain $\hat{m}$
\State Regress $(Y - \hat{g}(X, Z))^2$ on $X, Z$ using $D_2$ to obtain $\hat{v}$
\State Set $\hat{f}(x,z) \gets (\hat{g}(x, z) - \hat{m}(z))/\hat{v}(x,z)$
\State Regress $Y$ on $Z$ using $D_1$ to compute residuals $\hat\epsilon_i$
\State Regress $\hat{f}(X,Z)$ on $Z$ using $D_1$ to compute residuals $\hat\zeta_i$
\State Compute $T$ as in \eqref{eq:T-pcm} using $\hat\epsilon_i$ and $\hat\zeta_i$, 
$i \in \{1, \dots, n\}$
\State Compute $p$-value: $p \gets 1 - \Phi(T)$ 
\State \Return{$p$}
\end{algorithmic}
\end{algorithm}

Due to the sample splitting, the $p$-value of the PCM is a random quantity. We
can compute the PCM on several different splits to produce multiple $p$-values
that can be dealt with using standard corrections for multiple testing. In
practice, we follow the recommendation of the original paper and compute the
$p$-value as in step 9 of Algorithm~\ref{alg:pcm} but instead using the average
of the test statistics from the different splits. We denote the number of
different splits by $K$ and use 5--10 in the applications. The resulting test
should be conservative which results in a power loss, however, the test averaged
from different splits should still be more powerful than a single application of
the PCM due to more efficient use of the data. If one desires a perfectly
calibrated $p$-value from multiple splits, it is possible to use the method in
\citet{guo2023rank} but we do not pursue this further here.

\subsubsection{Comparison of the GCM and PCM tests}\label{sec:comparison}

The GCM and PCM tests not only differ in terms of their target quantities,
but also regarding computational aspects. 
The GCM test requires the regression of $Y$ on $Z$ and $X$ on 
$Z$. This prohibits the use of the GCM in settings where $X$ is a
high-dimensional or non-tabular data modality and can not
be represented as or reduced to a low-dimensional tabular modality. 
The PCM test, on the other hand, does not require regressing $X$ on 
$Z$. Thus, the PCM test allows the end-to-end
use of non-tabular data modalities, such as images or text, for instance,
via the use of deep neural networks. In contrast to the GCM, the PCM relies 
on sample splitting and requires more regressions and may thus be less
data-efficient. This is addressed, in parts, by repeating the PCM 
test with multiple random splits, as described above.

\subsection{Data sets}\label{sec:datasets}

\subsubsection{Variable significance testing: CCLE data}

We consider a subset of the anti-cancer drug dataset from the Cancer Cell Line
Encyclopedia \citep[CCLE,][]{barretina2012cancer} which contains the response to
the PLX4720 drug as a one-dimensional, continuous summary measure obtained from
a dose-response curve and a set of $1638$ mutations (absence/presence coded as
0/1, respectively) in $n = 472$ cancer cell lines. 
To obtain comparable results, we follow
the pre-processing
steps in \citet{bellot2019cond} and \citet{shi2021double} 
by screening
for mutations which are marginally correlated with drug response $S \coloneqq
\{j \in [1638] : \lvert\operatorname{Cor}(Y, X_j)\rvert > 0.05\}$, which leaves
$\lvert S\rvert = 466$ mutations.
See Section~\ref{sec:discussion:ccle} for a discussion of data-driven
pre-screening of mutations on type-I error control.

\subsubsection{Modality selection: TCGA data}

We consider the openly available TCGA HCC multiomics data set used in
\citet{chaudhary2018deep,poirion2021deepprog}. The preprocessed data consist of
survival times for $n = 360$ patients with liver cancer together with RNA-seq
($\RNA \in \RR^{15629}$), miRNA ($\miRNA \in \RR^{365}$), and DNA methylation
($\DNA \in \RR^{19883}$) modalities.
Pre-processing involved the removal of features and samples which contained more
than 20\% missing values and imputation of the remaining missing values. Further
detail can be found in \citet{chaudhary2018deep}. 

\subsubsection{Modality selection with imaging: MIMIC data}

We consider the MIMIC Chest X-Ray data set
\citep{johnson2019mimic,sellergren2022simplified}, which contains 
the race ($R$; with levels ``white'', ``black'', ``asian''), sex 
($X_1$; with levels ``male'', ``female''), age ($X_2$, in years), 
pre-trained embeddings of chest x-rays ($Z$) and (among other
response variables) whether a pleural effusion 
($Y$) was visible on the x-ray for $n = 181342$ patients.
The dimension of the image embedding was reduced by using the
first 111 components of a singular value decomposition, which 
explain 98\% of the variance.

\subsection{Computational details}

All analyses were carried out using the \proglang{R} language for statistical
computing \citep{pkg:base}. The covariance measure tests are implemented in
\pkg{comets} \citep{pkg:comets} which relies on \pkg{ranger} \citep{pkg:ranger}
and \pkg{glmnet} \citep{pkg:glmnet} for the random forest and LASSO regressions,
respectively. Code for reproducing all results is available at
\url{https://github.com/LucasKook/comets}. In the following, unless specified
otherwise, GCM and PCM tests are run with random forests for all regressions.
LASSO regressions are used for analyzing the TCGA data in
Section~\ref{sec:multi}.
A \proglang{Python} implementation of COMETs, the \pkg{pycomets} library
\citep{pkg:pycomets}, is available on GitHub
\url{https://github.com/shimenghuang/pycomets}.

\begin{table*}[!ht]
\centering
\caption{%
Results for the CCLE data in Section~\ref{sec:varsig}. The table shows variable
importance ranks and $p$-values for the relation of mutations of ten genes with
the response to PLX4720 conditional on the 465 other mutations in the data. The
PCM test was run with $K = 10$ random splits. The variable importance ranks
(obtained via random forests, RF, or elastic net regression, EN) and the CRT, 
GCIT and DGCIT results were obtained from \protect{\citet{bellot2019cond}} and 
\protect{\citet{shi2021double}}. 
}\label{tab:varsig}
\resizebox{\textwidth}{!}{%
\begin{tabular}{lrrrrrrrrrr}
\toprule
\bf Method & \multicolumn{10}{c}{\bf Gene mutations} \\ 
& BRAF\_V600E & BRAF\_MC & HIP1 & FLT3 & CDC42BPA & 
THBS3 & DNMT1 & PRKD1 & PIP5K1A & MAP3K5 \\ \midrule
EN & 1 & 3 & 4 & 5 & 7 & 8 & 9 & 10 & 19 & 78 \\
RF & 1 & 2 & 3 & 14 & 8 & 34 & 28 & 18 & 7 & 9  \\
\midrule
CRT & $<0.001$ & $<0.001$ & $0.008$ & 0.017 & 0.009 & 0.017 & 0.022 & 0.002 & 0.024 & 0.012\\
GCIT & $<0.001$ & $<0.001$ & $0.008$ & 0.521 & 0.050 & 0.013 & 0.020 & 0.002 & 0.001 & $< 0.001$\\
DGCIT  & $<0.001$ & $<0.001$ & $<0.001$ & $<0.001$ & $<0.001$ & $<0.001$ & $<0.001$ & $<0.001$ & $<0.001$ & 0.794 \\
\midrule
GCM & 0.030 & 0.033 & 0.010 & 0.005 & 0.004 & 0.042 & 0.010 & 0.165 & 0.464 & 0.504\\
PCM & 0.001 & 0.012 & 0.008 & 0.009 & 0.014 & 0.027 & 0.014 & 0.011 & 0.022 & 0.022\\
\midrule
GCM (no screening) & 0.003 & 0.021 & 0.006 & 0.002 & 0.002 & 0.068 & 0.007 & 0.007 & 0.223 & 0.216\\
PCM (no screening) & 0.002 & 0.007 & 0.082 & 0.151 & 0.186 & 0.134 & 0.138 & 0.108 & 0.198 & 0.122\\
\bottomrule
\end{tabular}}
\end{table*}

\section{Results}\label{sec:results}

With our analyses, we aim to show how testing with covariance measures can be
used to tackle two of the most common supervised learning problems in 
biomedical applications with multimodal data: Variable significance testing and
modality selection (see Figure~\ref{fig:overview}). Throughout, we compare 
COMETs with existing methods (if applicable) on openly available real
data sets (see Section~\ref{sec:datasets} for an overview of the data sets).

\subsection{Variable significance testing}\label{sec:varsig}

We apply COMETs to the anti-cancer drug dataset from the Cancer Cell Line
Encyclopedia \citep{barretina2012cancer} and compare with the results obtained
using the CRT \citep{candes2016panning} GCIT \citep{bellot2019cond} and DGCIT
\citep{shi2021double}.
See Section~\ref{sec:intro} for information on the CRT and Model-X based tests. 
The null hypotheses $H_0(j) : Y \indep X_j \given X_{-j}$ 
are tested for $j \in S$ to detect mutations that are significantly
associated with PLX4720 drug response. 

\subsubsection{COMETs identify mutations associated with PLX4720 drug activity}

Table~\ref{tab:varsig} summarizes the results for the GCIT, DGCIT, GCM, and
PCM test and the ten selected mutations in \citet[Figure~4]{bellot2019cond}.
Overall, there is large agreement between all tests which all reject the null 
hypothesis for the BRAF\_V600E, BRAF\_MC, HIP1, FLT3, THBS3 and DNMT1 mutations,
corroborating previously reported results. For the PRKD1, PIP5K1A and MAP3K5 
mutations, the PCM test rejects while the GCM test does not, which is consistent 
with the PCM test having power against a larger class of alternatives 
(Figure~\ref{fig:gcmpcm}).

\subsubsection{COMETs detect relevant mutations without pre-screening}

Prior results rely on pre-screening genes based on their marginal correlation
with the drug response. However, marginal correlation cannot inform subsequent
conditional independence tests in general and the data-driven pre-screening may
have lead to inflated false positive rates \citep{berk2013valid}. However, the 
GCM and PCM test can be applied without pre-screening and still consistently 
reject the null hypothesis of conditional independence for the BRAF\_V600E and
BRAF\_MC mutations (see rows in Table~\ref{tab:varsig} with ``no screening'').
When correcting (Holm) the $p$-values to attain a family-wise error rate
of $5\%$ for
the ten mutations of interest, the GCM and PCM still reject the null hypothesis
for BRAF\_V600E ($p = 0.024$ for the GCM test and $p = 0.020$ for the PCM test).
This rejection is expected because PLX4720 was designed as a BRAF inhibitor
\citep{barretina2012cancer}.

\subsection{Modality selection}\label{sec:multi}

The goal of our analysis is to identify modalities among RNA-seq, miRNA
and DNA methylation that are important for predicting survival of liver
cancer patients by testing if the event is independent of the modality 
$M_j$ given the other modalities $M_{-j}$, $j \in \{1, 2, 3\}$. This is 
a challenging problem due to the high dimensionality of both the candidate
modality $M_j$ and the conditioning variables in $M_{-j}$.

\subsubsection{Evidence of DNA methylation being important for 
predicting survival in liver cancer patients}

Table~\ref{tab:multi} (PCM-RF) shows $p$-values for the PCM test ($K=10$ different splits)
testing for significance of the RNA-seq, miRNA and DNA methylation modalities
conditional on the remaining two without pre-screening features in any of
the modalities using a random forest regression. There is some evidence that 
the DNA methylation modality is important for predicting death in liver cancer 
patients. Conversely, the PCM test 
does not provide evidence that survival depends on
the RNA-seq or miRNA modalities, when already conditioning on 
the DNA methylation data. Comparable results are obtained when substituting the
random forest regression for $Y$ on $\RNA$, $\miRNA$, and $\DNA$ with a 
cross-validated LASSO regression using the optimal tuning parameter: After a
multiple testing correction (Holm), both PCM tests reject the null hypothesis 
only for the DNA methylation modality.

\begin{table}[!ht]
    \centering
    \caption{%
    Results ($p$-values) for the multiomics application in 
    Section~\ref{sec:multi} using the PCM with $K=10$ random splits once using
    a random forest (RF) for the regression of $Y$ on $\RNA$, $\miRNA$ and $\DNA$,
    and once a cross-validated high-dimensional linear regression (LASSO).
    }\label{tab:multi}
    \resizebox{\tabwidth}{!}{%
    \begin{tabular}{lrr}
    \toprule
     \bf Null hypothesis & \bf PCM-RF & \bf PCM-LASSO\\
     \midrule
     $H_0 : Y \indep \RNA \given \miRNA, \DNA$ & 0.178 & 0.066 \\
     $H_0 : Y \indep \miRNA \given \RNA, \DNA$ & 0.165 & 0.044 \\
     $H_0 : Y \indep \DNA \given \RNA, \miRNA$ & 0.014 & 0.002 \\
     \bottomrule
    \end{tabular}}
\end{table}

\subsection{Modality selection with imaging data}\label{sec:mimic}

Using deep learning methods,
\citet{glocker2022risk} provide evidence that both the race and the response 
(pleural effusion) can be predicted from the x-ray embedding with high 
accuracy. The goal of our analysis is to test, whether race helps predict the
response when already conditioning on age, sex, and the x-rays and, vice versa,
whether the x-rays contain information for predicting pleural effusion given
sex, age, and race.

\subsubsection{Strong evidence for x-ray imaging and race being
important for predicting pleural effusion}

There is strong evidence against the null hypotheses of pleural effusion being
independent of either x-ray imaging or race given the other and, additionally, 
sex and age of a patient (Table~\ref{tab:mimic}). 

\begin{table}[!ht]
    \centering
    \caption{%
    Results ($-\log_{10}$-transformed $p$-values) for the GCM and PCM applied 
    to the full MIMIC data set in Section~\ref{sec:mimic}. Both tests reject
    both hypotheses. See Figure~\ref{fig:mimic} for an uncertainty assessment.
    }\label{tab:mimic}
    \begin{tabular}{lrr}
    \toprule
     \bf Null hypothesis & \bf GCM & \bf PCM \\
     \midrule
     $H_0 : Y \indep R \given X, Z$ & 6.158 & 77.762 \\
     $H_0 : Y \indep Z \given X, R$ & 13805.802 & 1270.361 \\
     \bottomrule
    \end{tabular}
\end{table}

To gauge the uncertainty in the results of the COMETs, we 
repeat the tests on 75 random (non-overlapping) subsamples of different sample
sizes (150, 600, 2400) of the data. Only the PCM rejects the null hypothesis of 
pleural effusion (PE) being independent of race given the x-ray, sex, and age 
of a patient at any of the considered sample sizes which provides evidence that 
$\E[\varepsilon \xi]$ is close to zero yet $\E[Y \given X,Z]$ still varies 
non-linearly with $X$. At full sample size, the GCM does reject, indicating the
presence of a weak linear signal (estimated correlations between pleural
effusion and race residuals are smaller than $0.015$). It is somewhat unsurprising
to see both COMETs reject the null hypothesis at such large sample sizes
\citep[Section~4.3]{greenland2019valid}. 

\begin{figure}[!ht]
    \centering
    \includegraphics[width=\figwidth]{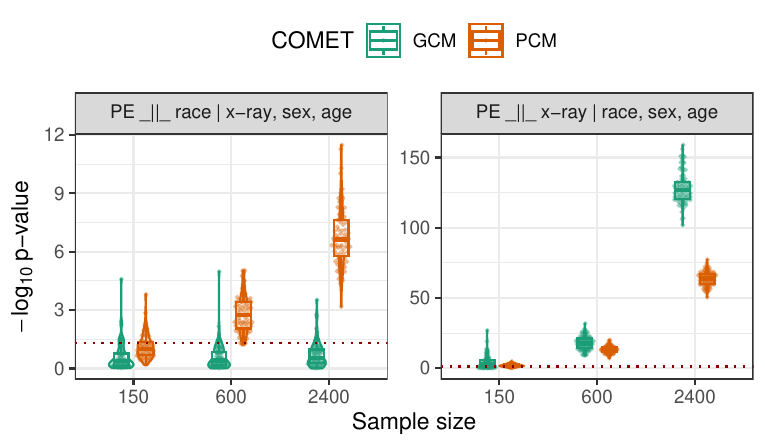}
    \caption{
    Results ($-\log_{10}$-transformed $p$-values) for the GCM and PCM applied to
    75 random non-overlapping splits of different sample sizes ($n \in\{150, 600,
    2400\}$) of the MIMIC data set in Section~\ref{sec:mimic}. Splitting the data 
    enables an analysis of the uncertainty in the tests' rejections and the strength of
    evidence against the null.
    }\label{fig:mimic}
\end{figure}

Both tests reject the null hypothesis of pleural effusion (PE) being independent
of x-ray given race, sex and age of a patient at any sample size but in fact the
GCM produces smaller $p$-values. This indicates that there is a significant 
component in $\E[Y \given X, Z]$ varying linearly with $X$; in these cases the
PCM will not outperform the GCM for a fixed sample size.

\subsection[Computation times]{Computation times}
\label{sec:comptimes}

\begin{figure}[!ht]
\centering
\includegraphics[width=\figwidth]{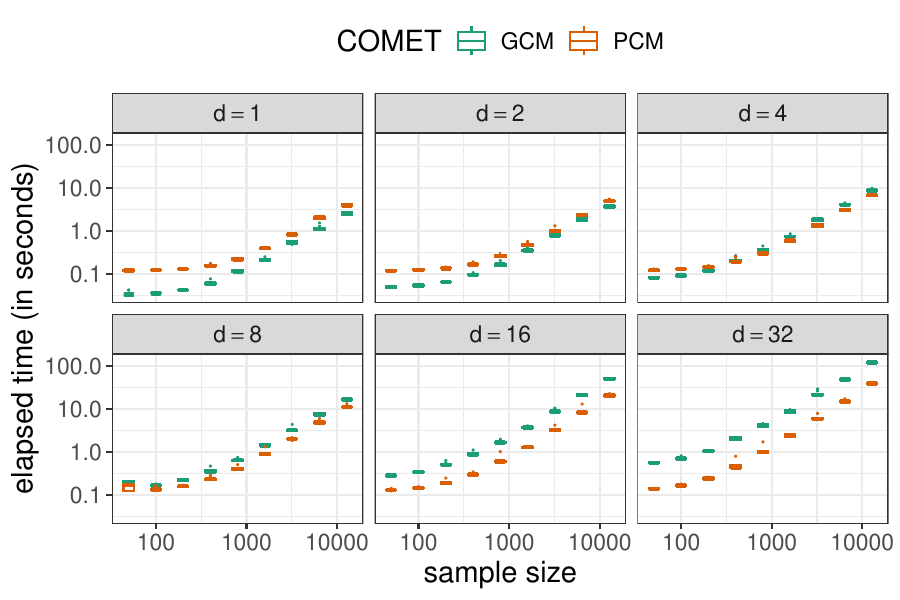}
\caption{%
Computation times (in seconds; y-axis) for the GCM and PCM test using random
forest regressions for varying dimensionality of $X$ (panels) and sample size
(x-axis).
}\label{fig:comptime}
\end{figure}

The computation time of the GCM and PCM test depends on the dimensionality $d$
of $X \in \RR^d$ and sample size $n$ and the chosen regression methods. For
low-dimensional $X$, the PCM test requires more regressions than the GCM test
which results in slower computation times (see Figure~\ref{fig:comptime}).
However, for higher-dimensional $X$, the GCM test requires more regressions
resulting in longer computation times. For moderate dimensions ($d = 4$ and $d =
8$), the computation times are similar.

\section{Discussion}\label{sec:discussion}

We present COMETs for algorithm-agnostic significance testing with multi-modal,
potentially non-tabular data, which relies on tests of conditional independence
based on covariance measures. The versatility of the GCM and PCM
tests is shown in several applications involving variable significance testing
and modality selection in the presence of high-dimensional conditioning 
variables. In the following, we discuss the applications in more detail and
end with a discussion of computational aspects and recommendations for using
COMETs in supervised learning applications with multimodal data.

\subsection{Variable significance testing}
\label{sec:discussion:ccle}

The GCM and PCM test show comparable results to competing methods and can be
applied without relying on data-driven pre-screening which otherwise can
invalidate $p$-values and lead to inflated type~I error rates. Type~I error
control additionally suffers from the performed number of tests. After 
correcting for multiple testing, the COMETs provide evidence
that BRAF\_V600E is associated with PLX4720 activity while controlling for
all other mutations. As highlighted before, this is expected since PLX4720
was designed as a BRAF inhibitor.

\subsection{Modality selection}

The PCM test is applied to the TCGA data set to test which modalities (RNAseq,
miRNA, DNAm) are important (conditional on the others) for predicting survival
in liver cancer patients and rejects the null hypothesis for the DNA methylation
modality. Failure to reject the null hypothesis for the RNA-seq and miRNA
modalities may be due to the low sample size and extremely high dimensionality
of the problem and ought to be interpreted as lack of evidence that RNA-seq and
miRNA data contain information for predicting survival beyond DNA methylation in
the data at hand. Taken together, this application demonstrates that the PCM
test can be used for modality selection with high-dimensional candidate and
conditioning modalities. COMETs could, for instance, be used to trade off the
economic cost of measuring an omics (or imaging, as in the MIMIC application)
modality with the gain in predictive power at a given significance level.
It is worth noting that a naive test based on the comparison of cross-validated
mean-squared errors using all variables and all but one variable does not result
in a valid statistical test
\citep{williamson2021nonparametric,lundborg2022projected}. 
Lastly, the validity of conditional independence tests applied to the TCGA data
depends on the validity of the imputation procedure used during data
pre-processing.

\subsection{Modality selection with imaging data}

The large and openly available MIMIC data set serves as an example application
of how image and other non-tabular modalities may enter an analysis based on
covariance measure tests. The PCM does not require pre-trained embeddings and
could, in principle, also be used in combination with deep convolutional neural
networks if the raw imaging data is available. The 111-dimensional embedding
further enables the use of the GCM test to serve as a benchmark. However, it is
important to properly choose the regressions involved in COMETs as the tests
rely on their quality and asymptotic properties
\citep{shah2020hardness,lundborg2022projected}. Nevertheless, to the best of our
knowledge, no other tests exist with theoretical guarantees that also permit
testing $Y \indep X \given Z$ when $X$ is a non-tabular modality.

\subsection{Recommendations and outlook}

As outlined in Section~\ref{sec:comparison}, the regression of $X$ on $Z$
required by the GCM can become computationally challenging if $X$ is
high-dimensional (which is why the GCM test is not applied in
Section~\ref{sec:multi} for modality selection) or non-tabular (this was
circumvented by using the relatively low-dimensional tabular embedding of the
chest x-ray images in Section~\ref{sec:mimic}; see also the
computation times in Section~\ref{sec:comptimes}). The PCM test, in contrast,
does not rely on this regression and is thus directly applicable in cases where
$X$ and $Z$ are high-dimensional or non-tabular modalities. The GCM has further
been adapted to settings with functional outcomes
\citep{lundborg2022conditional}, continuous time stochastic processes
\citet{christgau2023nonparametric}, censored outcomes \citep{kook2023model}, and
extended to powerful weighted \citep{scheidegger2022weighted} and kernel-based
\citep{fernandez2022general} versions. These are all COMETs proposed in the
literature and we leave their applicability in biomedical contexts as a topic
for future work.

In the applications presented in this paper, random forest and LASSO regressions
were used. Random forests are computationally fast and require little
hyperparameter tuning to obtain well-performing regression estimates. However,
for very high-dimensional applications in which the number of features exceeds
the number of observations, the LASSO is a fast and computationally stable
alternative.

Overall, we believe that COMETs provide a useful tool for bioinformaticians to
assess significance in applications with high-dimensional and potentially
non-tabular omics and biomedical data 
while appropriately controlling error probabilities.
The increasing familiarity of data analysts with supervised learning methods, on
which COMETs rely, help safeguard the validity of the statistical inference.
Further, the algorithm-agnostic nature of the procedures makes COMETs easily
adaptable to future developments in predictive modeling.

\section*{Acknowledgments}
We thank Niklas Pfister and David R\"ugamer for helpful discussions. We thank
Klemens Fr\"ohlich, Witold Wolski and Shimeng Huang for helfpful comments on the
manuscript. LK was supported by the Swiss National Science Foundation (grant no.
214457). ARL was supported by a research grant (0069071) from Novo Nordisk
Fonden.

\bibliographystyle{abbrvnat}
\bibliography{reference}

\end{document}